\documentclass[aps,prb,a4paper,10pt,twocolumn,floatfix,preprintnumbers,amsmath,amssymb,superscriptaddress,longbibliography]{revtex4-1}

\usepackage{float}
\usepackage{graphicx}
\usepackage{dcolumn}
\usepackage{bm}
\usepackage{color}
\usepackage{xspace}

\usepackage{textcomp} 

\graphicspath{{./}{figure/}}
\usepackage[colorlinks,plainpages=false,linkcolor=blue,urlcolor=blue,citecolor=blue,pdfpagemode=UseNone,pdfstartview=FitBH]{hyperref}
\usepackage{dcolumn,graphicx,color,booktabs,microtype,afterpage}
\usepackage[charter,greekuppercase=italicized]{mathdesign}

\newcommand{\tcr}[1]{\textcolor{black}{#1}}

\begin{document}

\title{Chemical and hydrostatic-pressure effects on the Kitaev honeycomb material Na$_2$IrO$_3$}

\author{G.\ Simutis}
\email{gediminas.simutis@psi.ch}
\affiliation{Laboratory for Muon Spin Spectroscopy, Paul Scherrer Institut, Villigen PSI, Switzerland}

\author{N.\ Barbero}
\email{nbarbero@phys.ethz.ch}
\affiliation{Laboratorium f\"ur Festk\"orperphysik, ETH Z\"urich, CH-8093 Z\"urich, Switzerland}

\author{K.\ Rolfs}
\affiliation{Laboratory for Developments and Methods, Paul Scherrer Institut, CH-5232 Villigen, Switzerland}

\author{P.\ Leroy-Calatayud}
\affiliation{Laboratorium f\"ur Festk\"orperphysik, ETH Z\"urich, CH-8093 Z\"urich, Switzerland}

\author{K.\ Mehlawat}
\affiliation{Indian Institute of Science Education and Research, Sector 81, Mohali 140306, India}

\author{R.\ Khasanov}
\affiliation{Laboratory for Muon Spin Spectroscopy, Paul Scherrer Institut, Villigen PSI, Switzerland}

\author{H.~Luetkens}
\affiliation{Laboratory for Muon Spin Spectroscopy, Paul Scherrer Institut, Villigen PSI, Switzerland}

\author{E.\ Pomjakushina}
\affiliation{Laboratory for Multiscale Materials Experiments, Paul Scherrer Institut, CH-5232 Villigen, Switzerland}

\author{Y.~Singh}
\affiliation{Indian Institute of Science Education and Research, Sector 81, Mohali 140306, India}

\author{H.-R.\ Ott}
\affiliation{Laboratorium f\"ur Festk\"orperphysik, ETH Z\"urich, CH-8093 Z\"urich, Switzerland}
\affiliation{Paul Scherrer Institut, CH-5232 Villigen PSI, Switzerland}

\author{J.\ Mesot}
\affiliation{Laboratorium f\"ur Festk\"orperphysik, ETH Z\"urich, CH-8093 Z\"urich, Switzerland}
\affiliation{Paul Scherrer Institut, CH-5232 Villigen PSI, Switzerland}

\author{A.\ Amato}
\affiliation{Laboratory for Muon Spin Spectroscopy, Paul Scherrer Institut, Villigen PSI, Switzerland}

\author{T.\ Shiroka}
\affiliation{Laboratorium f\"ur Festk\"orperphysik, ETH Z\"urich, CH-8093 Z\"urich, Switzerland}
\affiliation{Paul Scherrer Institut, CH-5232 Villigen PSI, Switzerland}

\date{\today}

\begin{abstract}
The low-temperature magnetic properties of \tcr{polycrystalline} Na$_2$IrO$_3$, a candidate material for the realization of a quantum spin-liquid state, were investigated by means of muon-spin relaxation and nuclear magnetic resonance methods under chemical and hydrostatic pressure. The Li-for-Na chemical substitution promotes an inhomogeneous magnetic order, whereas hydrostatic pressure (up to 3.9\,GPa) results in an enhancement of the ordering temperature $T_\mathrm{N}$.
In the first case, the inhomogeneous magnetic order suggests either short- or long-range correlations of broadly distributed $j=\,$\textonehalf\ Ir$^{4+}$ magnetic moments, reflecting local disorder. The increase of $T_\mathrm{N}$ under applied pressure points at an increased strength of three dimensional interactions arising from interlayer compression.
\end{abstract}

\maketitle{}

\section{Introduction}
\begin{figure}
\centering
\includegraphics[width={0.65\columnwidth}]{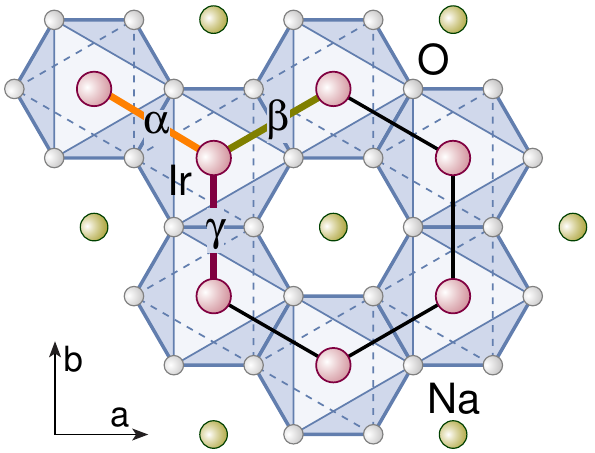}\\
\caption{\label{fig:honeycomb}%
Schematic sketch of the honeycomb layer in Na$_2$IrO$_3$. The effective spin-\textonehalf\ 5$d^5$ Ir$^{4+}$ ions
(at the center of oxygen octahedra)
interact via three distinct NN (nearest neighbor) exchange couplings,
here denoted by $\alpha$, $\beta$, and $\gamma$.
The $J$ and $K$ parameters in Eq.~(\ref{eq:KHhamiltonian})
represent the Heisenberg- and Kitaev exchange integrals, respectively,
taken over all the NN links in the honeycomb.
}
\end{figure}

Magnetic frustration resulting from bond-dependent exchange interactions is a possible route to the formation of a quantum spin-liquid (QSL) state,\cite{Balents2010} in which quantum effects prevent a long-range ordering of magnetic moments even at zero temperature. The Kitaev-Heisenberg model on a single-layer honeycomb lattice with bond-dependent Ising interactions (see Fig.~\ref{fig:honeycomb}), which
can be solved exactly in certain cases, is known to host QSL as one of the possible states.
The latter has been predicted to exhibit a range of unconventional features, such as emergent
Majorana fermions and gauge fluxes as effective excitations.\cite{Kitaev2006}

Real systems featuring the interactions assumed in the Kitaev model may be found in spin-orbit coupled materials with edge-shared octahedra.\cite{Jackeli2009} However, experiments have shown that only a few compounds exhibit the relevant Kitaev interactions
required to realize this model. The most prominent examples are the layered honeycomb-lattice iridates Na$_2$IrO$_3$\cite{Singh2010,Singh2012} and $\alpha$-Li$_2$IrO$_3$,\cite{Omalley2008,Chaloupka2010} as well as the recently identified H$_3$LiIr$_2$O$_6$.\cite{Kitagawa2018} In addition, a few three-dimensional systems that realize bond-frustrated lattices have been discovered, including $\beta$ and $\gamma$ polytypes of Li$_2$IrO$_3$.\cite{Takayama2015,Modic2014}
While all the above materials have bond-dependent anisotropic (i.e. Kitaev-type) interactions, also non-negligible Heisenberg and symmetric off-diagonal exchange interactions, resulting in competing
ground states, have to be taken into account. In particular, all the known candidate materials, except H$_3$LiIr$_2$O$_6$, adopt long-range ordered magnetic structures at low temperatures. Since the character of these states depends on the exact details of the underlying Hamiltonian, different materials exhibit distinct magnetic structures.

Considering that subtle modifications may lead to rather different ground states, it is intuitively tempting to deliberately induce variations of states or phases, e.g., by changing the chemical composition or by varying external parameters such as pressure. From this perspective, Na$_2$IrO$_3$ represents a particularly interesting case. Initial measurements have shown that substituting Na by Li leads to a substantial decrease of the magnetic ordering temperature $T_\mathrm{N}$.\cite{Cao2013} Later studies pointed out that,
at high levels of Li substitution, phase separation unfortunately prevents a wider range of tunability.\cite{Manni2014} Nevertheless, it was suggested  that even low-level substitution, well below the phase-separation threshold, may transform the original zigzag magnetic order into a spiral one.\cite{Rolfs2015}
Subsequently, optical-spectroscopy measurements showed that Li-substitution 
reduces the metal-metal hopping integral $t$ compared to the Coulomb repulsion term $U$,
thus enhancing the magnitude of the Mott insulating gap.\cite{Hermann2017}
Despite these interesting results, to date a microscopic investigation of the magnetic properties of Na$_2$IrO$_3$ under chemical or hydrostatic pressure is still missing.

In this report, we use muon-spin relaxation ($\mu$SR) and nuclear magnetic resonance (NMR) techniques as local probes and examine the evolution of magnetism in Na$_2$IrO$_3$ upon chemical doping (Li for Na substitution) and under applied hydrostatic pressure. In the former case, we find that at low substitution levels the magnetic order is rather robust, whereas above a $\sim 5\%$-substitution threshold, an inhomogeneous static magnetic order sets in. In case of applied pressure, up to 3.9\,GPa, we find a linear increase of the ordering temperature with pressure, but no qualitative changes in the magnetic ground state. The increased $T_\mathrm{N}$ indicates an enhancement of effective exchange interactions, most likely, arising from a reduction of the interlayer distances.\cite{Hermann2017}

\section{The Kitaev-Heisenberg model}
To date, Na$_2$IrO$_3$ is the only system to have been directly shown to host Kitaev-like interactions.\cite{Chun2015} To fully describe the interactions between the effective spin one-half Ir$^{4+}$ magnetic moments (see Fig.~\ref{fig:honeycomb}), a so-called \emph{extended} Kitaev-Heisenberg Hamiltonian on a honeycomb lattice is required:\cite{Katukuri2014}
\begin{equation}
\begin{split}
\mathcal{H} = \sum_{<ij>}^{}\sum_{\alpha, \beta, \gamma}^{}[J(p) \mathbf{S_i}\cdot\mathbf{S_j} + 2K(p) {S_i}^{\gamma}{S_j}^{\gamma} +\\\Gamma(p)({S_i}^{\alpha}{S_j}^{\beta} + {S_i}^{\beta}{S_j}^{\alpha})].
\label{eq:KHhamiltonian}
\end{split}
\end{equation}
The Heisenberg exchange coupling $J$ and the Ising-like Kitaev coupling
$K$ define the \emph{plain} Kitaev-Heisenberg model Hamiltonian.
Here the double summation runs over the bond- ($\alpha$, $\beta$, and
$\gamma$) and NN- ($i$, $j$) indexes, respectively.
Depending on the ratio between the $J$ and $K$ interaction parameters,
the zero-field ambient-pressure solution predicts six possible
ground states.\cite{Chaloupka2013}
These include two QSL states, a ferromagnetic phase, and three antiferromagnetic phases, the latter exhibiting three possible spin arrangements (N\'eel, stripy, and zigzag).\cite{Chaloupka2013} The \emph{extended} model, represented in Eq.~(\ref{eq:KHhamiltonian}), includes also a $\Gamma$ parameter, which captures the symmetric off-diagonal exchanges and requires a numerical solution. Finally, \tcr{since all the coupling parameters  depend on orbital hybridization, in turn depending on structural details,} a possible pressure dependence of the coupling parameters is included to adapt the Hamiltonian to our case.

\section{Experimental details}
Na$_{2-x}$Li$_x$IrO$_3$ powder samples with $x=0$, 0.05, 0.1, and 0.15 were synthesized following the procedure described in Ref.~\onlinecite{Rolfs2015} and used for the $\mu$SR measurements. The solid-state reaction synthesis used IrO$_2$, Na$_2$CO$_3$, and Li$_2$CO$_3$ as starting materials, mixed in stoichiometric ratios and heated up to 1000$^\circ$C. \tcr{The original characterization~\cite{Rolfs2015} showed
a lattice contraction upon Li doping, with Li uniformly replacing the Na ions. Bulk magnetic measurements indicated that both the Curie-Weiss and the transition temperature stays approximately the same, independent of Li concentration, in contrast with reports on single crystals.\cite{Cao2013,Manni2014}}
For the NMR measurements, polycrystalline Na$_2$IrO$_3$ samples were produced using
a similar solid-state synthesis, as reported in Ref.~\onlinecite{Singh2010}. A sample from this second batch, denoted as sample 2, was also used for the high-pressure $\mu$SR measurements (up to 2.4\,GPa).

Muon-spin relaxation ($\mu$SR) measurements were performed using the continuous muon beam at the Paul Scherrer Institute, Villigen, Switzerland. Compounds with different Li substitution values were studied using the low-background instrument GPS,\cite{Amato2017} with the powders placed in a fly-by fork-type sample holder. The high-pressure $\mu$SR measurements were performed at the GPD instrument,\cite{Khasanov2016} where the samples were placed into a double-wall pressure cell, similar to those described in Refs.~\onlinecite{Khasanov2016,Shermadini2017}. The pressure was transmitted to the sample via Daphne oil 7373. Zero-pressure experiments in the pressure cell were used to cross-calibrate the absolute magnetic volume fractions by comparing the measurements with the results from GPS experiments. Data analysis was performed using the \texttt{musrfit} program.\cite{Suter2012}

In addition to $\mu$SR, we also performed complementary high-pressure NMR measurements. By employing a hybrid piston-clamped zirconia-anvil cell,\cite{Barbero2018} smaller samples could be probed via NMR at higher pressures (up to 3.9\,GPa). The NMR investigations under pressure included line-shape and spin-lattice $T_1$ relaxation-time measurements in an applied magnetic field of 7.057\,T. The inner part of the  pressure cell was filled with Daphne oil 7575, acting as a pressure-transmitting medium. The applied hydrostatic pressure was monitored via the pressure-dependent NQR (nuclear quadrupole resonance) signal of ${}^{63}$Cu in Cu$_2$O.\cite{Reyes1992} The most suitable nucleus for our study was $^{23}$Na, an $I=3/2$ nucleus with a 100\% abundance and 79.47\,MHz Larmor frequency in the chosen applied magnetic field.

\section{Experimental Results}
\subsection{Weak transverse-field $\mu$SR measurements}
\begin{figure}
\centering
\includegraphics[width={0.8\columnwidth}]{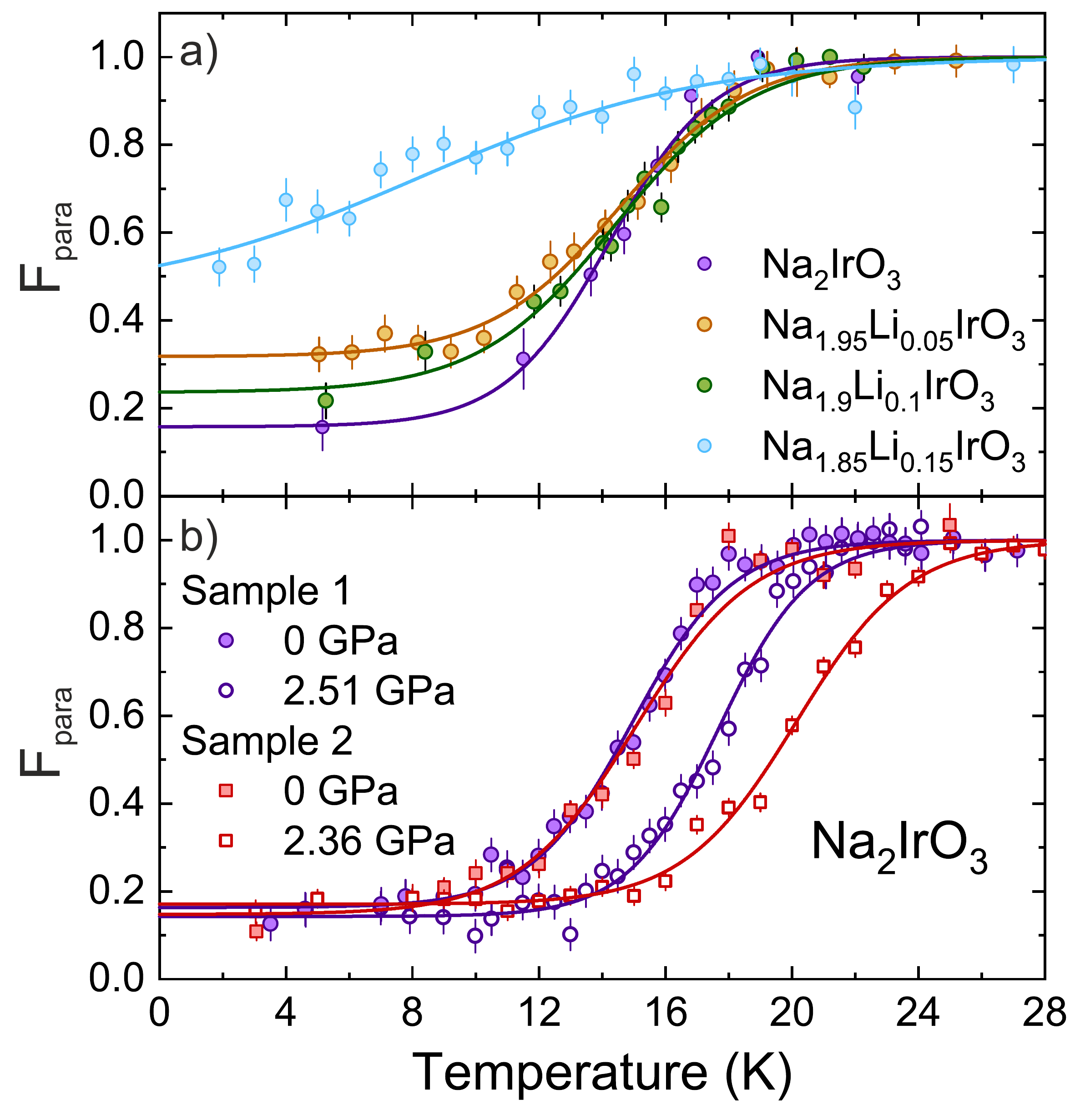}\\
\caption{\label{fig:wTF_GPSD}
Paramagnetic fraction as a function of temperature for samples with
different Li-substitution levels (a) and for Na$_{2}$IrO$_{3}$
measured at different pressures (b),
as obtained from the $\mu$SR experiments.
Lines are fits using Eq.~(\ref{eq:erf}).
}
\end{figure}

Weak\footnote{Unless the muon spin is made to rotate, the use of weak magnetic fields is required in order to avoid significant deflections of the muon beam.} transverse-field (wTF) $\mu$SR can be used to determine the magnetic ordering temperature and the magnetic volume fraction. In the paramagnetic phase muon spins precess coherently around the externally applied field. Upon cooling, the electronic magnetic moments order and the resulting magnetic-field distribution at the muon stopping sites dephases the muon spins. The ensuing damped oscillation can be fitted with a simple harmonic model and the resulting metadata serve to determine the paramagnetic volume fraction $F_\mathrm{para}$. The latter is related to the parameters of a magnetic-phase transition via:\cite{Khasanov2008}
\begin{eqnarray} \label{eq:erf}
F_\mathrm{para}(T) = F_\mathrm{nm} + (1-F_\mathrm{nm})/\left[\exp\left(\frac{T_\mathrm{N}-T}{\Delta}\right) + 1 \right].
\end{eqnarray}
Here $F_{\mathrm{nm}}$ is a residual fraction, corresponding to that part of the sample which is not magnetically ordered even at the lowest temperatures, $T_\mathrm{N}$ is the inflection point of each $F_\mathrm{para}(T)$ curve, interpreted as the transition temperature, and $\Delta$ is the broadening parameter.\footnote{The transition width $\Delta T$, defined as the distance in
temperature from the 10\% to the 90\% of the increase in $F_\mathrm{para}(T)$, can be calculated from the broadening parameter
$\Delta$, by using $\Delta T = 2\ln(9)\Delta$. An alternative way to parameterize the transition width is to employ the error function, as defined, for example, in Ref.~\onlinecite{Shiroka2011}. The two parameters are approximately related by $\Delta_\mathrm{erf} = 1.7\Delta $ } The evolution of the paramagnetic volume-fraction with temperature is shown in Fig.~\ref{fig:wTF_GPSD}, both as a function of Li content (a) and as a function of hydrostatic pressure (b).

\begin{figure}
\centering
\includegraphics[width={0.8\columnwidth}]{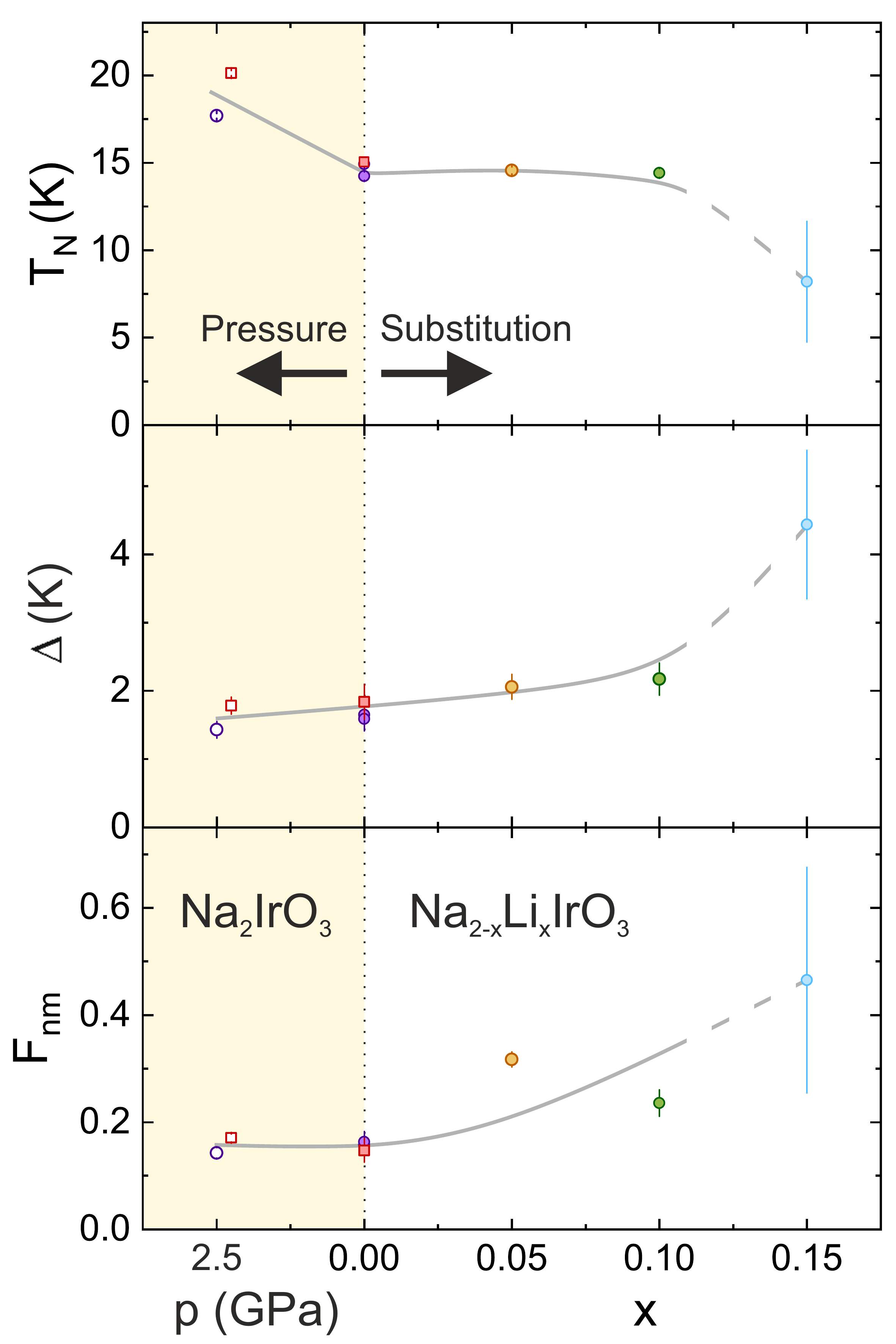}
\caption{\label{fig:wTF_vals}Best-fit values as extracted from fits of wTF-$\mu$SR data
with Eq.~(\ref{eq:erf}) for different pressures (left) and Li-substitution levels (right). Symbols
correspond to the measurements reported in Fig.~\ref{fig:wTF_GPSD}. Lines are guides to the eye.
}
\end{figure}

The best-fit parameters using Eq.~(\ref{eq:erf}) are summarized in Fig.~\ref{fig:wTF_vals}. While at low Li-substitution levels ($x = 0.05$ and 0.10) the transition temperature and the magnetically-frozen sample-fraction remain virtually unchanged, for $x = 0.15$ the transition
\tcr{is strongly broadened and the transition temperature drops significantly}. Hence, above a certain threshold, the disorder induced by chemical substitution \tcr{seems} to suppress the Ir$^{4+}$-based magnetic order, possibly making it short-ranged. \tcr{This is in contrast to earlier reports of a continuous reduction of the ordering temperature\cite{Cao2013,Manni2014} and may be related to the differences between single crystal and polycrystalline samples, as explained in the Discussion.}
On the other hand, the application of hydrostatic pressure induces an increase of the transition temperature and exhibits a narrower transition width, as indicated by the smaller $\Delta$ value compared to that of the Li-substituted case.

As shown in Fig.~\ref{fig:wTF_GPSD}b, even nominally pristine samples exhibit a broad transition, $\Delta \sim$ 2\,K and a nonzero residual fraction, $F_\mathrm{nm}(0) \sim 0.15$. These two effects are, most likely,
\tcr{due to the presence of stacking faults which are common in layered honeycomb materials, as well as to a small amount of an impurity phase}, whose properties were studied in Ref.~\onlinecite{Krizan2014}. We cover this point in detail in the Discussion section.

\subsection{Zero-field $\mu$SR measurements}
\begin{figure}
\centering
\includegraphics[width={0.8\columnwidth}]{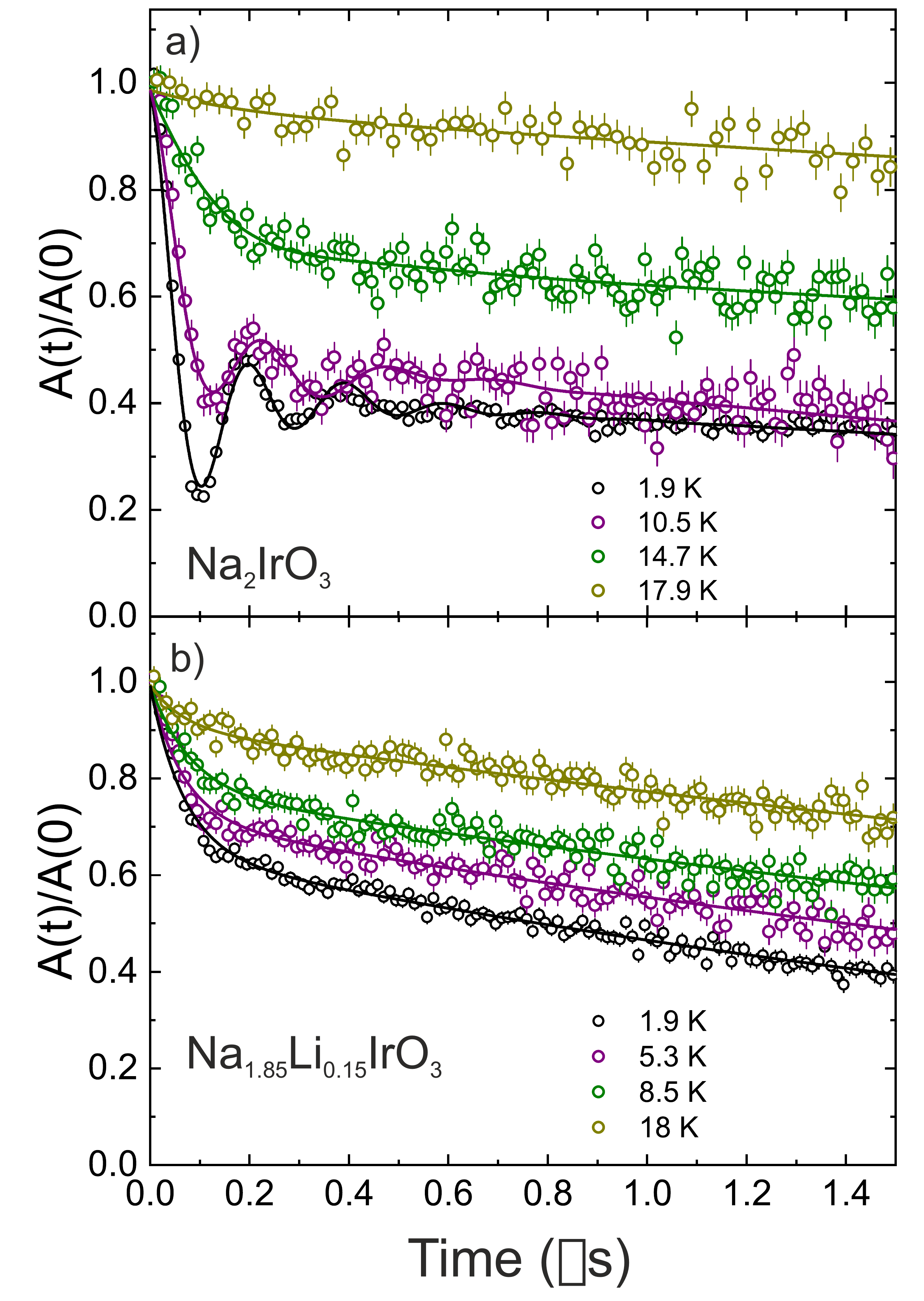}
\caption{\label{fig:ZF_spectra}Time-domain muon-decay asymmetries
for various temperatures as observed in Na$_2$IrO$_3$ (a) and in
Na$_{1.85}$Li$_{0.15}$IrO$_3$ (b). Note the lack of oscillations
in the second case.
}
\end{figure}

To get further insight into the magnetically-ordered phase of Na$_{2}$IrO$_{3}$, we performed zero-field (ZF) $\mu$SR measurements. Muon-decay asymmetry spectra of the pristine and the $x = 0.15$ compounds at selected temperatures are shown in Fig.~\ref{fig:ZF_spectra}. At high temperatures (above $T_\mathrm{N}$), muon spins retain their initial polarization for a long time. The small relaxation observed in this case is due to nuclear magnetic moments and weak dynamic effects. Upon cooling, well defined oscillations appear in the pure Na$_2$IrO$_3$ sample, as well as in those  with low Li concentrations ($x \leq 0.1$), indicating the onset of a long-range magnetic order. On the other hand, no oscillations show up for the sample with maximum disorder, $x = 0.15$. Yet, in this case, the initial fast relaxation rates suggest the presence of a static magnetism. Indeed, additional longitudinal-field (LF)-$\mu$SR measurements (see Fig.~\ref{fig:LF} in the Appendix) show a prompt recovery of the main part of the asymmetry, hence confirming that magnetic moments are frozen (i.e., behave as static on the $\mu$SR time scale).

The time evolution of the muon-decay asymmetry for the samples showing oscillations ($x \leq 0.1$) can be described by:\cite{Yaouanc2011}
\begin{eqnarray}
A(t)/A(0) = F_\mathrm{osc}[ F_1 \cos(\gamma_\mu B_1 t)\exp(-\lambda_1 t)  +  \nonumber \\
  (1-F_1)\cos(\gamma_\mu B_2 t) \exp(-\lambda_2 t)  ] +  \nonumber \\
  (1 - F_\mathrm{osc}) \exp(-\lambda_T t) \mathrm{,}\label{eq:polar-pure}
\end{eqnarray}
where $F_\mathrm{osc}$ is the total fraction of the oscillating signal, best described as the sum of two oscillating components with weights $F_1$ and $(1-F_1)$, related to two different muon sites; $(1 - F_\mathrm{osc})$ is the relaxing-only component, $B_1$ and $B_2$ are the local fields experienced by the implanted muons, whereas $\lambda_1$, $\lambda_2$, and $\lambda_T$ are the relaxation rates of the two oscillating- and one non-oscillating component, respectively. The ratio of the signals from the two muon-stopping sites was found to be temperature independent, yet it turned out to depend on the Li substitution level. Thus, in the $x = 0$ case $F_1$ was found to be 0.53(4), whereas in the Li-substituted compounds $F_1$ was 0.38(3) for $x = 0.05$ and 0.2(3) for $x = 0.1$.

On the other hand, the muon-decay asymmetry of the $x = 0.15$ sample, not showing oscillations, is described by the sum of two relaxing components:
\begin{equation}
\label{polar-15}
A(t)/A(0) = F_\mathrm{fast}\exp(-\lambda_\mathrm{fast} t) +
(1 - F_\mathrm{fast}) \exp(-\lambda_T t) \mathrm{,}
\end{equation}
where $F_\mathrm{fast}$ corresponds to the fast-relaxing part of the signal,  with the rest $(1 - F_\mathrm{fast})$ relaxing at a slower pace. \tcr{Interestingly, the slow-relaxing part of the asymmetry persists even in applied LF fields. This suggests persistent spin fluctuations,
likely arising from frustration and competition of different ground states, coexisting with the frozen state.}

\begin{figure}
\centering
\includegraphics[width={\columnwidth}]{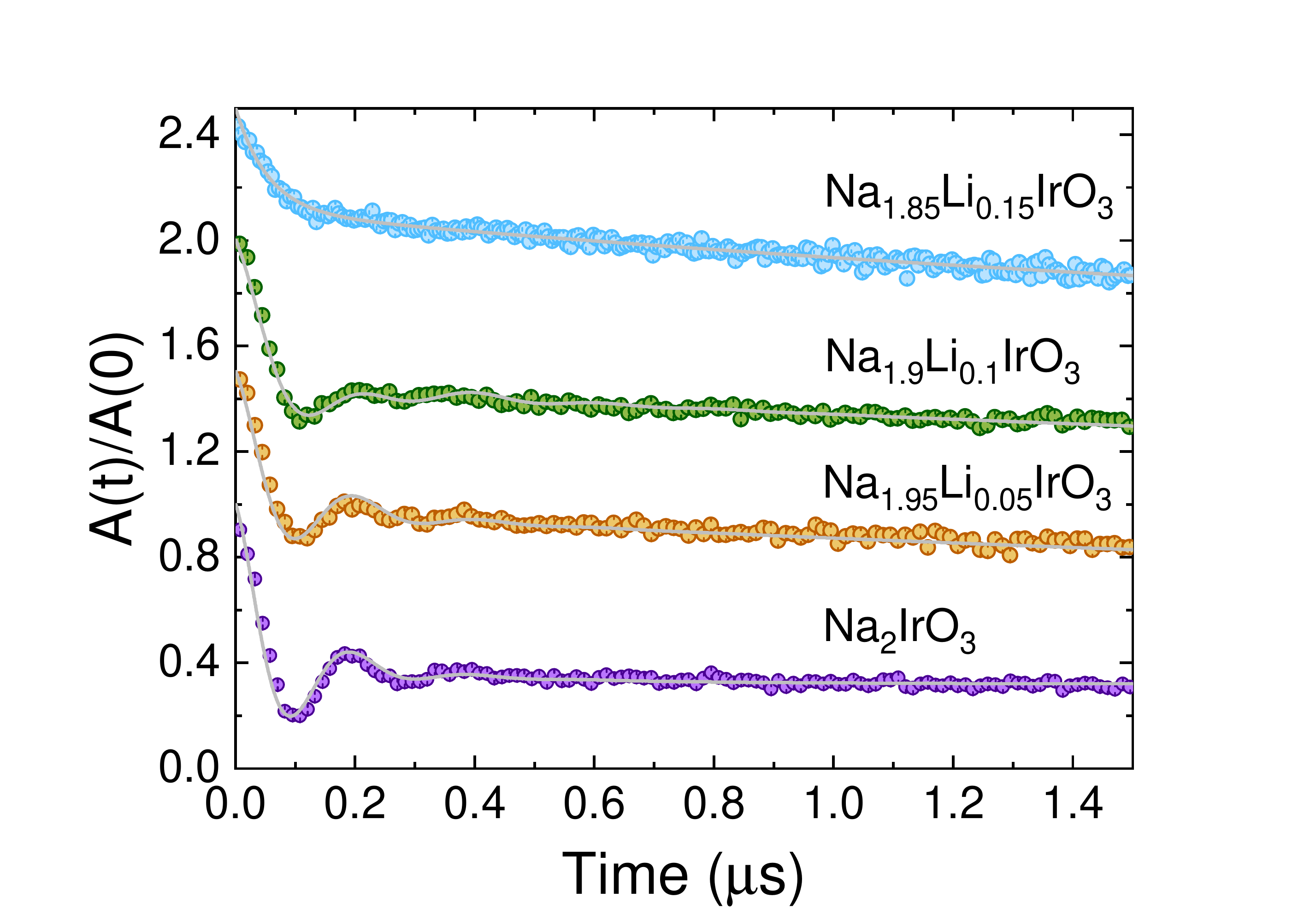}
\caption{\label{fig:ZF_LT}Muon-decay asymmetry as a function of time at 2\,K in the Li-substituted Na$_2$IrO$_3$ series. For clarity, the datasets are vertically offset by 0.5 units.}
\end{figure}

\begin{figure}
\centering
\includegraphics[width={0.8\columnwidth}]{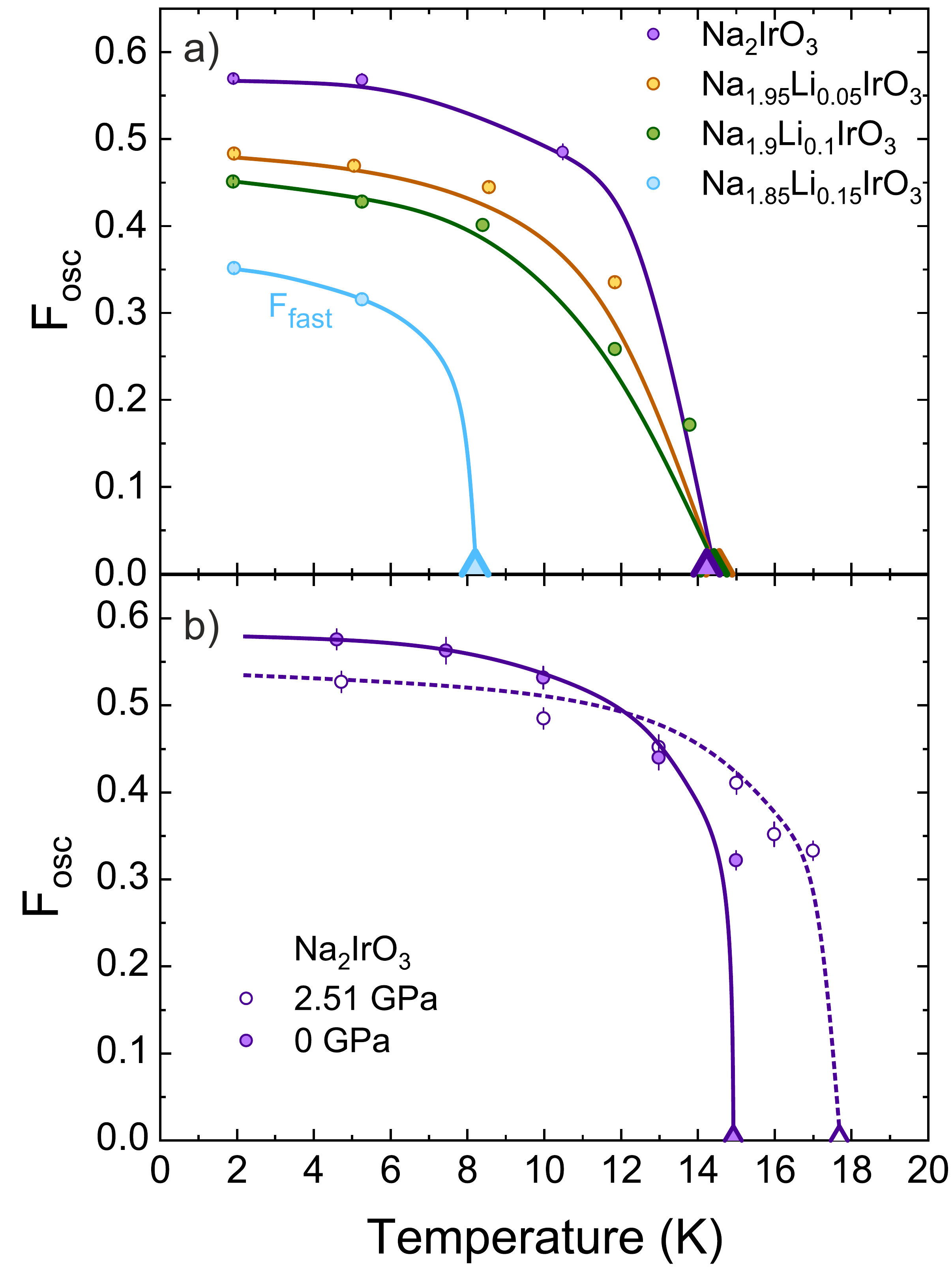}
\caption{\label{fig:ZF_par}Oscillating asymmetry $F_\mathrm{osc}$
as obtained from fits using Eq.~(\ref{eq:polar-pure}). In case of Li substitution we observe a clear decrease with increasing $x$ (a). \tcr{Since no oscillations are observed in the $x = 0.15$ case, the fast relaxing component of the asymmetry, $F_\mathrm{fast}$, is plotted, which corresponds to the static part of the sample.}
Oscillating asymmetry vs.\ temperature for Na$_2$IrO$_3$, measured at
ambient pressure and at $p = 2.51$\,GPa (b). The triangles indicate
the transition temperatures as obtained from transverse-field measurements.
Lines are guides to the eye.}
\end{figure}

\begin{figure}
\centering
\includegraphics[width={0.8\columnwidth}]{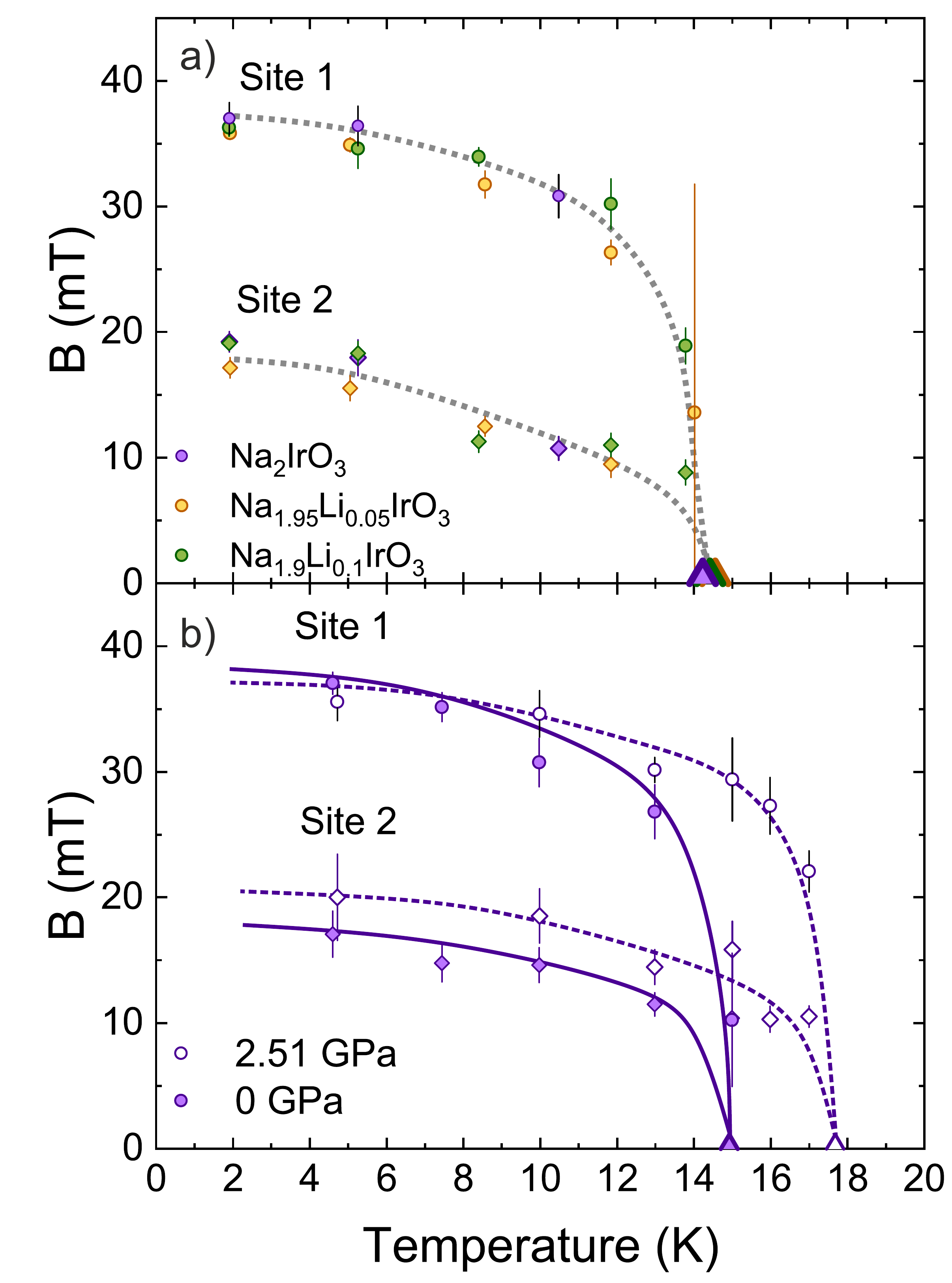}
\caption{\label{fig:ZF_parB}
Temperature dependence of the internal magnetic fields vs.\ temperature
for the Li-substituted samples where oscillations could be clearly
identified (a) and as a function of applied pressure for the pure
Na$_2$IrO$_3$ (b). Triangles denote the transition temperatures as
obtained from the weak transverse field measurements. Lines are
guides to the eye.}
\label{fig:fields_vs_T}
\end{figure}

Figure~\ref{fig:ZF_LT} shows an overview of the base-temperature (2\,K) spectra for all the samples. The well-defined oscillations observed in the $x = 0$ case are gradually suppressed as $x$ increases and disappear completely for $x = 0.15$. As can be seen from the extracted asymmetry parameters displayed in Fig.~\ref{fig:ZF_par}(a), the key change upon Li substitution is a monotonous reduction of the oscillating-signal fraction. At the same time, the internal field values, reported in Fig.~\ref{fig:ZF_parB}(a), remain virtually unchanged with substitution, showing only the expected reduction with increasing temperature.

A similar analysis was performed in the case of applied pressure. The resulting oscillating-asymmetry fraction is shown in Fig.~\ref{fig:ZF_par}(b). Unlike in the case of Li substitution, upon increasing pressure we observe only a small overall reduction of the oscillating asymmetry. Also the population of the two muon sites does not change much, from $F_1 = 0.58(5)$ measured in the cell at ambient pressure to 0.45(6) at 2.51\,GPa. Similarly, the saturation-field value remains unchanged under applied pressure but, as expected, the local fields persist to higher temperatures, as shown in Fig.~\ref{fig:ZF_parB}(b). The data confirm the clear enhancement of the transition temperature $T_\mathrm{N}$ under applied pressure, already identified in the weak transverse-field experiments and supported by the relevant NMR data presented below.

\subsection{NMR measurements}
\begin{figure}[ht!]
\includegraphics[width={0.8\columnwidth}]{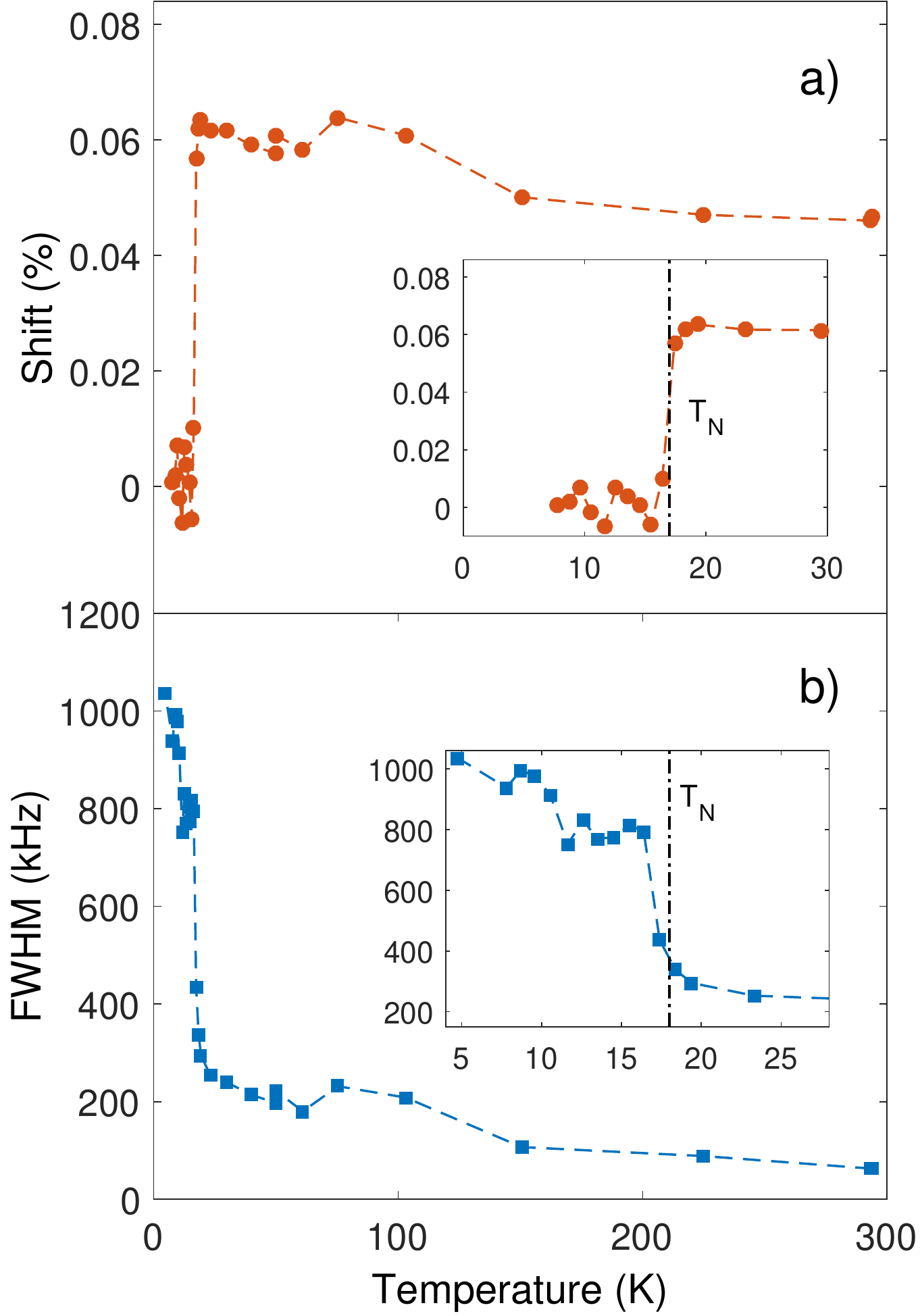}
\caption{\label{fig:NMR_shift_fwhm}Shifts (a) and FWHM widths (b) of the
${}^{23}$Na NMR lines in Na$_2$IrO$_3$, measured at 7.057\,T
and $p = 0$\,GPa from 4 to 295\,K. Insets highlight the drop in
shift and the increase in
line width occurring at $T_\mathrm{N}$. Uncertainties are of the order of the marker size.}
\end{figure}

The crystal structure of Na$_2$IrO$_3$ is monoclinic with space group $C12/m1$,\cite{Choi2012} where the spin-\textonehalf\ Ir$^{4+}$ ions are arranged on a honeycomb lattice (see Fig.~\ref{fig:honeycomb}). This results in ${}^{23}$Na NMR line-shapes consisting of the convolution of multiple lines, hence reflecting the three inequivalent Na-sites, as known from the crystal structure. As shown in Fig.~\ref{fig:NMR_lines} (in the Appendix), this complexity is further enhanced upon cooling.
At the onset of the (zigzag) antiferromagnetic (AFM) order in Na$_2$IrO$_3$,\cite{Chun2015} the ${}^{23}$Na line exhibits a sudden shift of $\sim +600$\,ppm [see Fig.~\ref{fig:NMR_shift_fwhm}(a)], suggesting the appearance of a spontaneous sublattice magnetization and corresponding magnetic field. At each temperature, the shift was defined as the relative deviation of $f_\mathrm{m}$ from the ${}^{23}$Na Larmor frequency (79.47\,MHz) in the applied magnetic field, with $f_\mathrm{m}$ the median NMR-spectrum frequency, sampled from 78 to 81\,MHz. Upon entering the magnetically ordered phase, lines also broaden significantly, as shown in Fig.~\ref{fig:NMR_shift_fwhm}(b). Their full width at half maximum (FWHM) was tracked over the entire covered temperature range. The steep increase at $T_\mathrm{N}$, typical of AFM transitions, signals the ordering temperature. To estimate the FWHM values, each spectrum
was integrated numerically in the relevant (78--81\,MHz) frequency interval.

\begin{figure}[ht]
\includegraphics[width={0.8\columnwidth}]{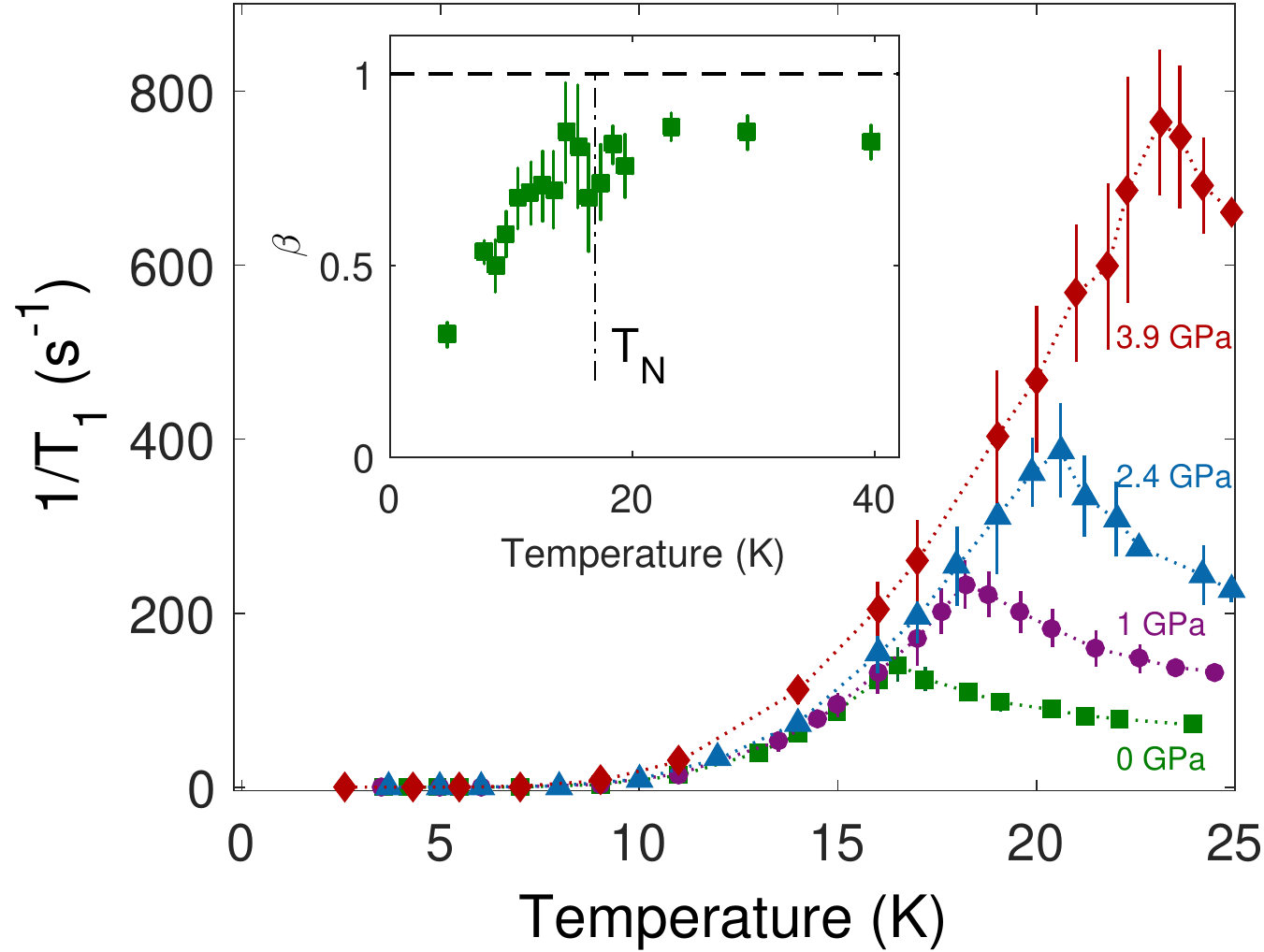}
\caption{\label{fig:NMR_HP}Na$_2$IrO$_3$ $1/T_1(T)$ relaxation rates from 4 to 25\,K, measured at the central-transition of the ${}^{23}$Na NMR lines in 7.057\,T, at ambient- and at three applied hydrostatic pressures (1, 2.4, and 3.9\,GPa). Inset: $\beta(T)$ variation across $T_\mathrm{N}$,
as resulting from ambient-pressure $T_1$ measurements.}
\end{figure}

To detect the onset of magnetic order under applied pressure, a faster and more accurate
way is to track the peak in the $1/T_1$ spin-lattice relaxation rate vs.\ temperature plot. To this end, the nuclear spin-lattice relaxation times $T_1$ were measured on resonance by means of the inversion-recovery method, using a spin-echo detection at variable delays. The $T_1$ values were determined by fitting a relaxation function relevant for spin-3/2 nuclei to the inversion-recovery data using:\cite{Mcdowell1995}
\begin{equation}
M_z(t)/M_0 = 1 - a\,[0.9\cdot e^{-(6t/T_1)^{\beta}}+0.1 \cdot e^{-(t/T_1)^{\beta}}].
\label{eq:T1_fit}
\end{equation}
Here $M_0$ is the saturation value of the nuclear magnetization, $a$ is an amplitude parameter (ideally 2), while the stretching coefficient $\beta$ accounts for the distribution of the spin-lattice relaxation times around a characteristic value $T_1$ ($\beta = 1$ for a single, well-defined spin-lattice relaxation rate $1/T_1$; $\beta < 1$ for an inhomogeneous distribution of $1/T_1$ values). At ambient pressure, NMR lineshape- and spin-lattice-relaxation results are in good agreement, both identifying the same $T_\mathrm{N}$ value, 16.4\,K. Such  value, taken as a reference,
is indicated by vertical dashed lines in Figs.~\ref{fig:NMR_shift_fwhm},\,\ref{fig:NMR_HP}, and \ref{fig:SQUID}.
Our $\mu$SR measurements on sample 1 instead indicate a $T_\mathrm{N}$ of $\sim$15\,K, compatible with previous studies\cite{Singh2010,Singh2012} and with our magnetization measurements. The latter were used to continuously check the sample
quality during all the experiments reported here (see Fig.~\ref{fig:SQUID} in the Appendix). Such a discrepancy suggests that the physical properties of Na$_2$IrO$_3$ depend substantially on the synthesis protocol, the handling procedure, and on the mosaicity of the crystal plane orientations, as discussed in detail in the Appendix.

\begin{figure}[ht]
\includegraphics[width={0.8\columnwidth}]{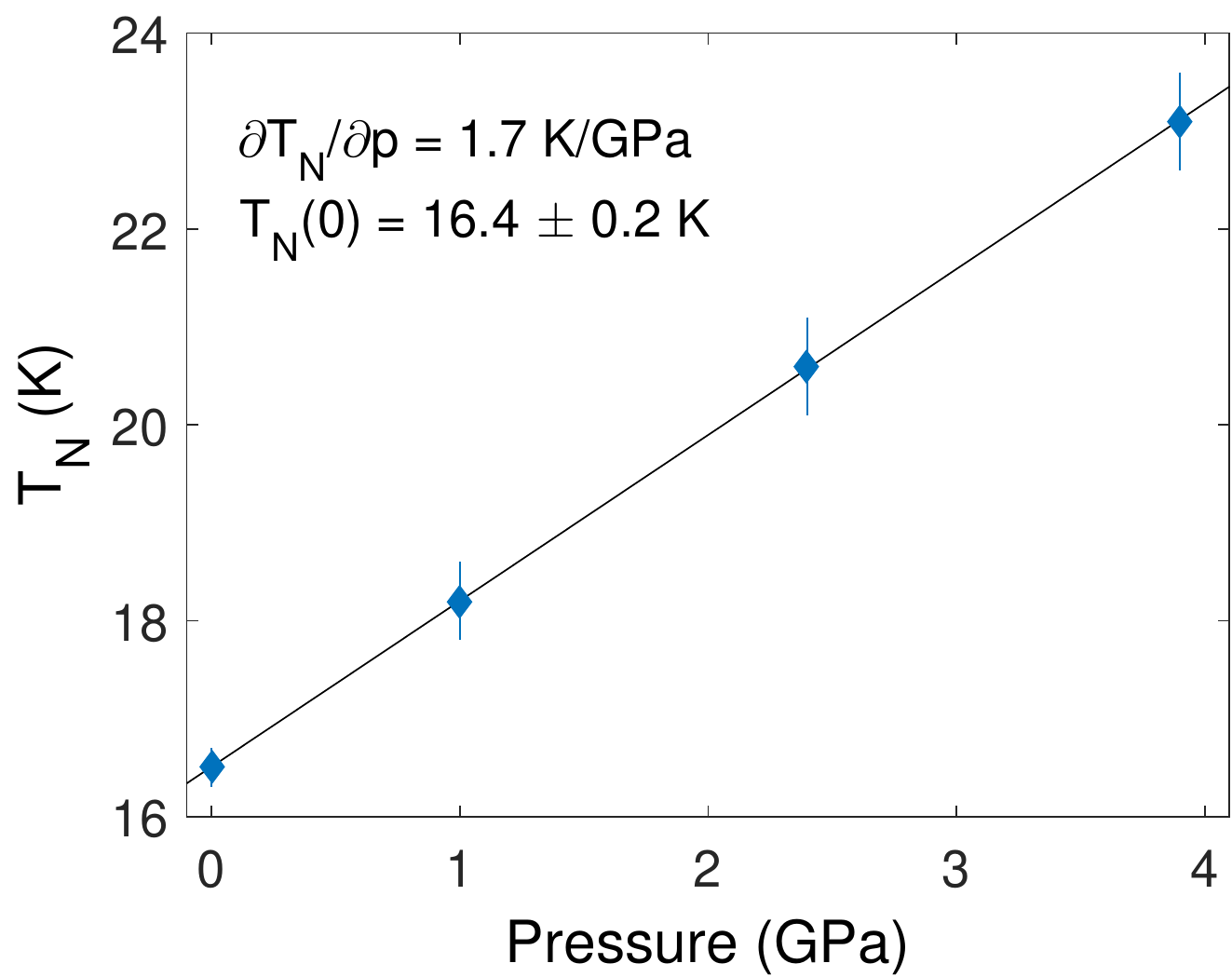}
\caption{\label{fig:NMR_values}$T_\mathrm{N}$ variation upon increasing
pressure. The gradient and intercept value, as determined from a
straight-line fit, are shown.}
\end{figure}

The $1/T_1$ values measured at different applied pressures are shown in Fig.~\ref{fig:NMR_HP}. A clear, well-defined cusp persists up to the highest pressures. The temperature values corresponding to the relaxation maxima are plotted in Fig.~\ref{fig:NMR_values}. This shows that the magnetic ordering temperature $T_\mathrm{N}$ increases linearly within the explored pressure range (up to 3.9\,GPa).

Below $T_\mathrm{N}$, the stretching parameter $\beta(T)$ --- used to fit the $T_1$ inversion-recovery curves --- shows a significant drop, indicative
of a broader distribution of $T_1$ values (see inset in Fig.~\ref{fig:NMR_HP}). Above $T_\mathrm{N}$ and up to room temperature, instead, $\beta(T)$ is constant with a value of $\sim$\,0.85. This indicates a narrow variance of $\sim$\,15\% in the distribution of the relaxation rates,\cite{Shiroka2011bis} most likely, related to tiny differences between the electronic environments probed by the three inequivalent ${}^{23}$Na sites.
The drop of $\beta(T)$ below $T_\mathrm{N}$ implies that the three inequivalent ${}^{23}$Na sites experience increasingly different relaxation rates, reflecting the enhanced inhomogeneity of fields and electronic environments in the AFM phase. At the same time, we found that pressure does not appreciably modify the stretching parameter $\beta$ and, therefore, does not affect significantly the $T_1$ distribution below the respective transition temperatures.

\section{Discussion}
In \tcr{polycrystalline Na$_2$IrO$_3$ samples}, the Li-for-Na substitution has a prominent effect on the ZF-$\mu$SR spectra which, upon a growing Li content, exhibit a gradual transition from well-defined asymmetry oscillations to spectra dominated by pure relaxation. Such behavior indicates a magnetic order which becomes increasingly \emph{inhomogeneous}. Since muons populate the local electrostatic minima throughout the sample volume, the measured spectra represent a convolution of signals arising from different parts of the system. As such, muon-spin asymmetry is directly related to the spatial distribution of magnetic moments and, in the present case, point to local disorder.

A similar asymmetry behavior has been observed also in other systems, including spin chains with bond- and site disorder,\cite{Thede2014b,Simutis2016}
or in iron-based superconductors at intermediate F doping.\cite{Shiroka2011} In all these cases, the high sensitivity of $\mu$SR to chemical modifications emphasizes the delicate nature of the (originally) homogeneous magnetic order which, nevertheless, does not evolve to a different type of magnetic structure. In fact, also our high-pressure $\mu$SR measurements on Na$_2$IrO$_3$ reveal that
the nature of the magnetic ground state remains virtually the same,
although the magnetic ordering temperature increases significantly
with pressure, at a rate of 1.6\,K/GPa.\footnote{This is an average value of measurements on two samples from different batches.}

The NMR measurements under hydrostatic pressure (up to 3.9\,GPa)
confirm the enhancement of $T_\mathrm{N}$ at an observed rate of 1.7\,K/GPa.
These results suggest the absence of pressure-induced phase transitions
within the explored pressure range. This conclusion is in good agreement with
results of high-pressure (up to 8\,GPa) optical-spectroscopy
and synchrotron x-ray diffraction measurements on Na$_2$IrO$_3$
single crystals, reported in Ref.~\onlinecite{Hermann2017}.
The same study also established that, in Na$_2$IrO$_3$,
the preferential compressibility along the $c$-axis direction
tends to reduce the distance between the honeycomb layers.\cite{Hermann2017}
This finding provides an intuitive explanation for our observation ---
namely, that the onset of magnetic order ultimately is favored
by the appearance of three-dimensional exchange interactions.
By analogy with a continuous-symmetry spin configuration,
as postulated by Mermin and Wagner,\cite{Mermin1966} 
a purely two-dimensional system cannot spontaneously break
the symmetry, i.e., it cannot achieve a magnetic order at $T>0$.
Nevertheless, in the Kitaev-Heisenberg model on a honeycomb lattice, due to anisotropy, the spin degrees
of freedom posses only a discrete symmetry. Here, the Mermin-Wagner theorem
provides only a qualitative framework,
since rigorously it can be applied only to cases of continuous rotation
symmetry, i.e., to the Heisenberg model.

It is worthwhile to compare the present results with the recently discovered
suppression of magnetic order in $\beta$-Li$_2$IrO$_3$,\cite{Majumder2018}
 occurring at an applied pressure of 1.4\,GPa. Before the
 vanishing of its magnetically-ordered state,
$\beta$-Li$_2$IrO$_3$ exhibits an intriguing behavior. While the ordered moments
maintain their magnitude
and the ordering temperature increases moderately (0.7\,K/GPa), the magnetic volume fraction drops drastically
upon applying even moderate pressures.
By contrast,  in Na$_2$IrO$_3$ we do not find a significant
reduction of the magnetic volume fraction with increasing pressure. This corroborates the former
statement, i.e.,
in our case, hydrostatic pressure essentially reduces
the distance between the honeycomb layers, whereas in $\beta$-Li$_2$IrO$_3$
(a three-dimensional Kitaev system), the whole hierarchy of
exchange interactions is drastically modified.

\tcr{Another closely related system is $\alpha$-Li$_2$IrO$_3$. At ambient pressure it has the same structure as Na$_2$IrO$_3$, but it was
shown that at about 3.8\,GPa, iridium ions dimerize.\cite{Hermann2018} Theoretical calculations in the same study suggested that the structural change is accompanied by a collapse of magnetic order. In general, there is a tendency of such systems to dimerization,\cite{Hermann2018,Biesner2018,Bastien2018,Veiga2017} but the characteristic pressure for Na$_2$IrO$_3$ may be much higher. Indeed, ab initio calculations reported in Ref.~\onlinecite{Hermann2018} corroborate this point of view, suggesting that in Na$_2$IrO$_3$ the dimerization might occur at 45\,GPa.
Such pressure is too high for the competing dimer-state to be relevant in our case.}

\tcr{Surprisingly, we found that Li-substituted polycrystalline samples do not follow the same trend as single-crystal samples.\cite{Cao2013,Manni2014} Thus, there is no continuous decrease in the ordering temperature; instead, the magnetic order becomes progressively more inhomogeneous and the transition is significantly broadened when the Li concentration reaches $x=0.15$. Such behavior may reflect 
the fine details of Li-substitution. Previous reports on single-crystal samples indicate a single preferential Li site,\cite{Manni2014}  whereas in our polycrystalline samples Li seems to replace all the Na sites with an equal probability.\cite{Rolfs2015} This difference may lead to different magnetic properties and indicates the difficulties in controlling disorder, especially in polycrystalline samples.}

The reported measurements also indicate the high sensitivity of Na$_2$IrO$_3$ to factors, such as the synthesis protocol and the handling procedure. The first requires a careful optimization of the solid-state reaction and of the
annealing protocol, so as to minimize the presence of
spurious phases, \emph{qualitatively} different from the pure one.
As for the handling procedure, this mainly involved carrying out experiments in an inert atmosphere.
Neglecting this precaution is known to affect the phase purity of Na$_2$IrO$_3$, implying
deteriorated samples over time, with a 
reduced magnetic susceptibility $\chi(T)$ and 
anomalous features below $T_\mathrm{N}$, as confirmed by our
time-dependent magnetization- (see Fig.~\ref{fig:SQUID}) and X-ray
scattering measurements.\cite{Krizan2014}
Finally, since crystal-growth protocols affect the physical
properties of the sample, they may induce $T_\mathrm{N}$ variations
ranging from 12 to 15\,K.\cite{Chun2015} In our case, high-pressure $\mu$SR measurements
of samples 1 and 2, belonging to different batches, indicate only slightly different transition temperatures.

In order to minimize the above issues, extra precautions were taken. For instance, the loading of the $\mu$SR pressure cell was performed in helium atmosphere in a glovebox, whereas the sealing of the NMR cell took place under argon flow.
Such measures were important, since preliminary measurements in air
resulted (a posteriori) in degraded samples.
In the Appendix we report comparative magnetic-susceptibility
measurements in pure and in degraded samples. Nevertheless, despite the above concerns, the tiny but non-negligible presence of an altered phase does not have any effect in the reported results since, as local-probe techniques, both $\mu$SR and NMR are site sensitive.

\section{Conclusion}
By using local magnetic probes, such as muon-spin relaxation and nuclear
magnetic resonance, we investigated the magnetic ground state of the
Na$_2$IrO$_3$ honeycomb iridate under hydrostatic pressure and in
case of Li-substitution. The chemical substitution of Na by Li
shifts the system
from a fully-ordered state towards inhomogeneous magnetic order.
Such inhomogeneous order suggests either short-range correlations or
long-range correlations of broadly distributed magnetic moments,
thus emphasizing the sensitivity to local disorder.
On the other hand, the application of hydrostatic pressure
is shown to enhance the ordering temperature, yet without
modifying the character of the magnetic ground state.
The increased $T_\mathrm{N}$ reflects
the preferential compressibility of Na$_2$IrO$_3$ along its
$c$-axis. This implies a reduced interlayer distance under pressure,
hence, a more pronounced 3D character, ultimately resulting in
the observed enhancement of transition temperature.

Our work confirms the challenges encountered in tuning the ground state of candidate Kitaev materials. Not only do we provide evidence about the sensitivity of honeycomb iridate Na$_2$IrO$_3$ to isovalent substitution and to hydrostatic pressure, but we also show how air-sensitivity and substitution-induced disorder may clearly affect the onset of antiferromagnetism in this compound. Even in pristine Na$_2$IrO$_3$, despite extensive
precautions in both the synthesis protocol and in the handling procedures, a tiny presence of impurities
cannot be excluded.

No evidence of quantum spin-liquid behavior was observed in the honeycomb iridate Na$_2$IrO$_3$ upon isovalent substitution or hydrostatic pressure. Taking into account the fragility of its magnetic state,
extensive experimental evidence will be required in order to unambiguously identify a
possible QSL state, considering that a lack of long-range order may also
be due to disorder or deterioration.

\section{Acknowledgements}
We would like to thank J.\ Barker for the technical assistance and D.\ Cheptiakov and L.\ Korosec for valuable discussions. Part of this work is based on experiments performed at the Swiss Muon Source S$\mu$S, Paul Scherrer Institute, Villigen, Switzerland. This work was financially supported in part by the Schwei\-ze\-rische Na\-ti\-o\-nal\-fonds zur F\"{o}r\-de\-rung der Wis\-sen\-schaft\-lich\-en
For\-schung (SNF). K.M.\ acknowledges the University Grants Commission, CSIR, India. Y.S.\ acknowledges the Ramanujan grants No.\ SR/S2/RJN-76/2010 and SB/S2/CMP-001/2013 from DST, India. K.R.\ acknowledges SNF Sinergia Project ''Mott physics beyond Heisenberg model''. Swiss National Science foundation has supported the work of G.S. (Grants No. 200021-149486 and No. 200021-175935) and N.B. (Grant No. 200021-169455). G.S.\ and N.B.\ contributed equally to this work.

\section*{Appendix}

\subsection{Lattice constants of Na$_{2-x}$Li$_{x}$IrO$_3$}
\tcr{Previously it has been shown that in single crystals of Na$_{2-x}$Li$_{x}$IrO$_3$ the lattice shrinks progressively upon Li for Na substitution, with the distances within the layers shortening faster.\cite{Manni2014} Our polycrystalline samples, characterized via neutron diffraction,\cite{Rolfs2015} exhibit the same shrinking effects. Therefore, Li substitution induces an effective chemical pressure, with a magnitude comparable to the single-crystal case}.\cite{Manni2014}
\begin{figure}[htb]
\centering
\includegraphics[width=0.8\columnwidth]{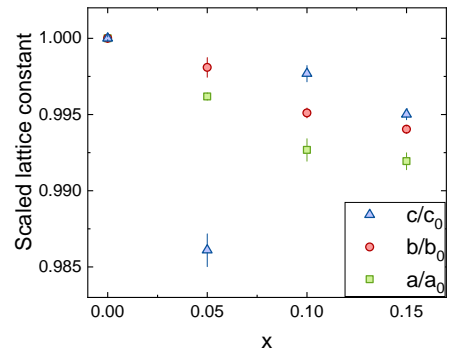}
\caption{\label{fig:Latt} Normalized lattice constants for different Li substitution levels.
The lattice constants for $x=0$ were taken from Ref.~\onlinecite{Manni2014}.}
\end{figure}

\subsection{Longitudinal-field $\mu$SR measurement of Na$_{1.85}$Li$_{0.15}$IrO$_3$}
\begin{figure}[htb]
\centering
\includegraphics[width=0.8\columnwidth]{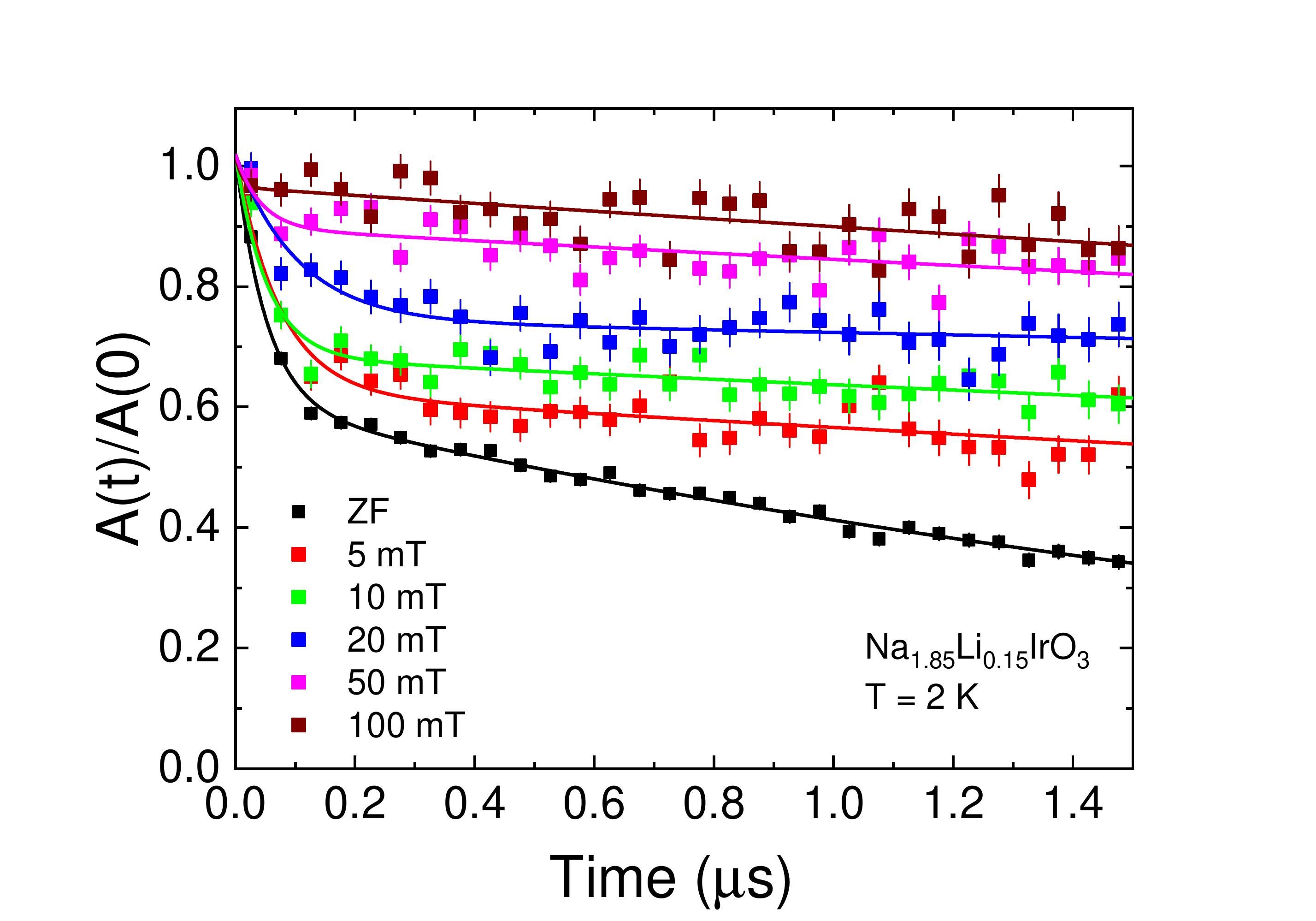}
\caption{\label{fig:LF}The decoupling of muon spins in an
applied longitudinal field indicates static magnetic moments
in Na$_{1.85}$\-Li$_{0.15}$\-IrO$_3$.
To highlight the recovery of asymmetry, a
data binning of 50\,ns was chosen, exceeding those in
the rest of the figures.}
\end{figure}

As reported above, the zero-field measurements of Na$_{1.85}$\-Li$_{0.15}$\-IrO$_3$, the
sample with maximum disorder, did not reveal oscillations in the muon-decay asymmetry,
but only a depolarization as a function of time. In principle, this could be due either to
fluctuating moments or to a static but inhomogeneous magnetic order. A good way to
differentiate between the two is to apply a longitudinal magnetic field. In case of static magnetic moments, the applied field decouples the muon spins, hence
recovering the decay asymmetry to its initial value. As shown in Fig.~\ref{fig:LF}
this is indeed the case for the $x = 0.15$ sample, where fields above 50\,mT fully
recover the asymmetry.
\subsection{$^{23}$Na NMR lines of pure Na$_2$IrO$_3$}
\begin{figure}[b!]
\includegraphics[width=0.8\linewidth]{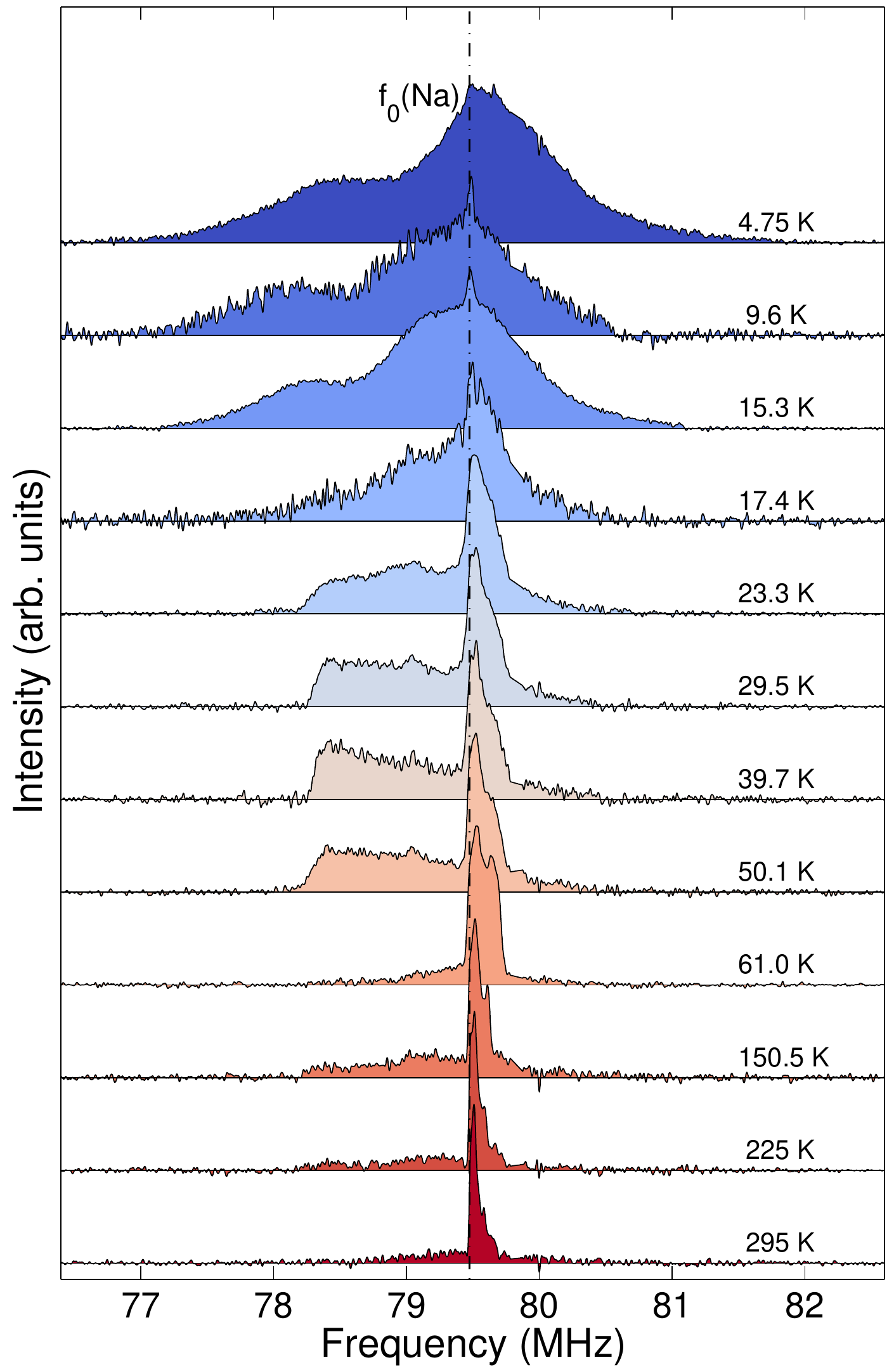}
\caption{\label{fig:NMR_lines} ${}^{23}$Na NMR lines in Na$_2$IrO$_3$ from 4 to 295\,K, measured at 7.057\,T. The vertical dashed line indicates the
Larmor frequency.}
\end{figure}
The $^{23}$Na NMR investigations of pure Na$_2$IrO$_3$ in applied pressure included line-shape and spin-lattice $T_1$ relaxation time measurements in a magnetic field of 7.057\,T. Typical NMR spectra,
as reported in Fig.~\ref{fig:NMR_lines}, were obtained via fast
Fourier transform of the spin-echo signal generated by $\pi/2$--$\pi$
rf pulses of 5 and 10\,$\mu$s and echo delays of 50\,$\mu$s.
The recycle delays ranged from 0.2\,s at room temperature up to 5\,s at 3\,K. As described in detail in the experimental-results
section, the three inequivalent Na sites exhibit different
dynamics and probe different electronic environments. This is
also reflected in an increased line complexity upon cooling.
Here the resulting line-shape is a complex convolution of
spectra from the three sites, each with different relaxation
times and electric-field gradients.
\subsection{Sample degradation issues and variations of $T_\mathrm{N}$}
\begin{figure}[ht]
\includegraphics[width={0.8\columnwidth}]{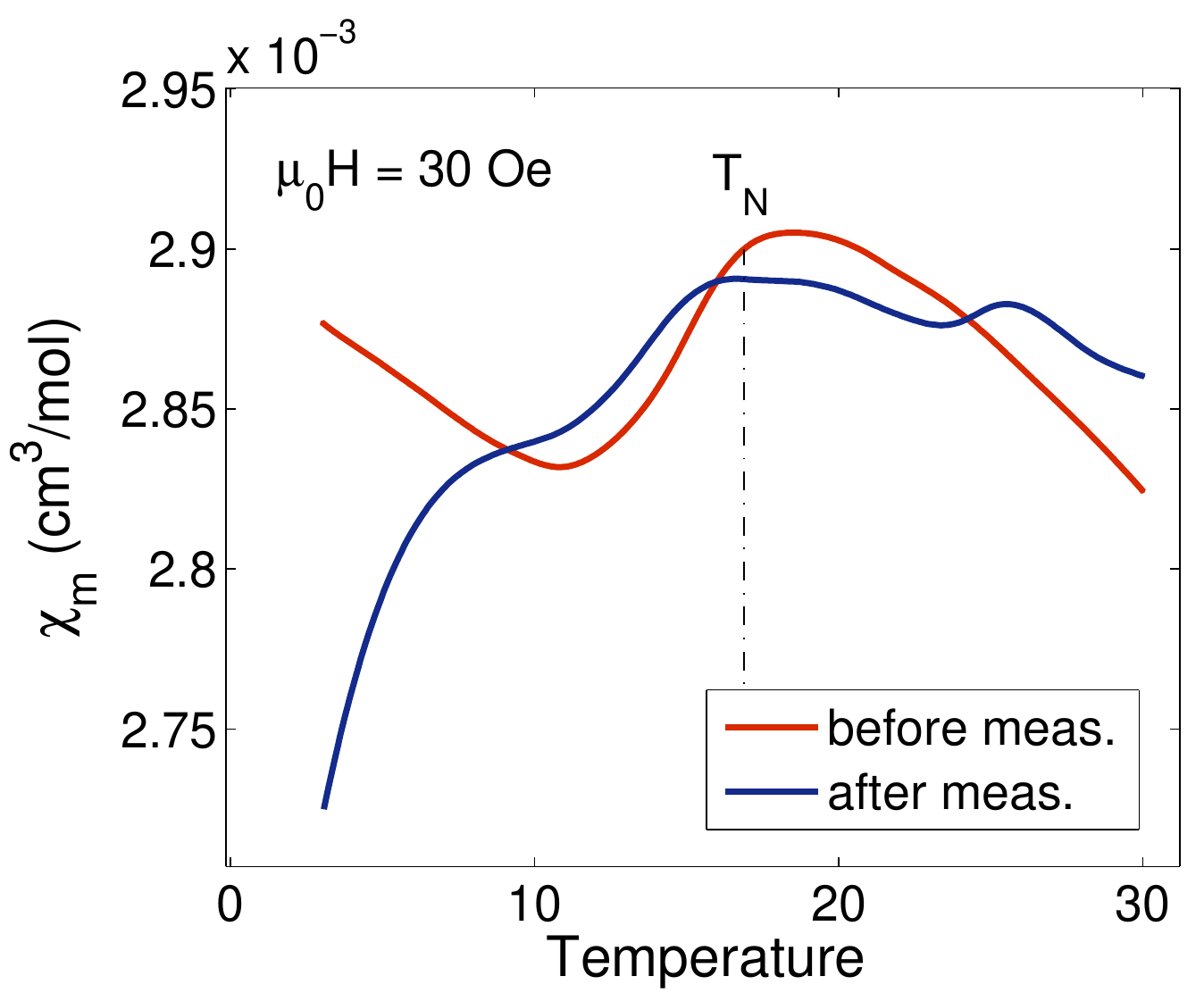}
\caption{\label{fig:SQUID}Magnetization measurements
of Na$_2$IrO$_3$ in an applied field of 3\,mT. The molar magnetic susceptibility
vs.\ temperature exhibits a clear change before (red line) and after
the preliminary NMR measurements (blue line), reflecting a sample degradation during handling.
This required a constant sample handling under Ar flow. As explained in the literature,\cite{deJongh1990}
the onset of magnetic order occurs at the inflection point between
the minimum and the maximum, here at $\sim$15\,K. The dotted vertical
line shows the $T_\mathrm{N}$ value as obtained from NMR data.
Further details about the uncertainty in identifying $T_\mathrm{N}$
are given in the discussion section.}
\end{figure}

%
The magnetization $M(T, H)$ measurements on Na$_2$IrO$_3$ were carried out
using a commercial MPMS XL-7 (magnetic property measurement system) with an
RSO (reciprocating sample option) in fields from 3\,mT to 7\,T, by covering a
temperature range from 3 to 300\,K. The $M(T, H)$ data were used to:
(a) check the sample quality before and after each measurement at ambient- or hydrostatic pressure and
(b) confirm that applied magnetic fields (up to 7\,T) do not significantly affect the onset of AFM, as previously reported.\cite{Ye2012}
Figure~\ref{fig:SQUID} reports the molar susceptibility $\chi_m(T) = M(T)/H$ of
Na$_2$IrO$_3$ across the AFM transition, measured at 3\,mT, before and after
preliminary NMR measurements. The data confirm that handling the sample without Ar flow and due diligence causes a significant sample degradation.

From the analysis of $\chi_m(T)$ data at 3\,mT, we find $\mu_{\mathrm{eff}} = 1.89$\,$\mu_{\mathrm{B}}$. The corresponding
Curie-Weiss temperature $\theta_{\mathrm{CW}}$ is $-122$\,K, in good agreement with previously reported values.\cite{Singh2012}
Since Ir$^{4+}$ ions exhibit a low-spin configuration, where $5d$ electrons populate only the $t_{2g}$ levels, crystal-field effects along with a non-negligible spin-orbit
coupling\cite{Singh2010} explain the higher effective magnetic moment $\mu_{\mathrm{eff}}$ with respect to
the spin-only value $\mu_{\mathrm{spin}} = 1.73$\,$\mu_{\mathrm{B}}$, predicted by theory. Finally, it is worth mentioning that
the presence of domains in the sample and its mosaicity affect
significantly the value of $T_\mathrm{N}$ with deviations up to 20\% from
the average value of 15\,K.\cite{Chun2015}

\bibliography{Na2IrO3bib}

\end{document}